\documentclass{ws-ijmpd}
\begin{document}

\title{Asymptotic freedom and quarks confinement treated through Thompson's approach}

\author{Cl\'audio Nassif*, A. C. Amaro de Faria** Jr. and P. R. Silva***}
\address{*UFOP: Universidade Federal de Ouro Preto, Campus Morro do Cruzeiro, s/n, CEP:34.500-000, Ouro Preto-MG, Brazil.\\
 **IEAv: Instituto de Estudos Avan\c{c}ados, Rodovia dos Tamoios, Km 099, CEP:12.220-000, S\~ao Jos\'e dos Campos-SP, Brazil.\\
 ***UFMG: Universidade Federal de Minas Gerais, Departamento de F\'{\i}sica-ICEx, Caixa Postal 702, CEP:30.123-970, Belo Horizonte-MG,
 Brazil.}

\par

\date{\today}
\maketitle
\begin{abstract}

In this work, we first use Thompson's renormalization group method to treat QCD-vacuum behavior close to the
regime of asymptotic freedom. QCD-vacuum behaves effectively like a ``paramagnetic system" of a classical theory
in the sense that virtual color charges (gluons) emerge in it as spin effect of a paramagnetic material when a
magnetic field aligns their microscopic magnetic dipoles. Making a classical analogy with the
paramagnetism of Landau's theory, we are able to introduce a kind of Landau effective action without temperature and
phase transition for just representing QCD-vacuum behavior at higher energies as being magnetization of a paramagnetic
material in the presence of a magnetic field $H$. This reasoning allows us to use Thompson's heuristic approach in
order to obtain an ``effective susceptibility" ($\chi>0$) of the QCD-vacuum. It depends on logarithmic of energy scale
$u$ to investigate hadronic matter. Consequently, we are able to get an ``effective magnetic permeability" ($\mu>1$)
of such a ``paramagnetic vacuum". As QCD-vacuum must obey Lorentz invariance, the attainment of $\mu>1$ must simply
require that the ``effective electrical permissivity" is $\epsilon<1$ in such a way that $\mu\epsilon=1$ ($c^2=1$).
This leads to the anti-screening effect, where the asymptotic freedom takes place. On the other hand, quarks
confinement, a subject which is not treatable by perturbative calculations, is worked by the present approach. We apply
the method to study this issue for obtaining the string constant, which is in agreement with the experiments.

\end{abstract}

\section{Introduction}

\hspace{3em}

The most largely employed strategy for dealing with problems involving many length scales is the ``Renormalization -
Group (RG) approach"\cite{1}. The RG has been applied to treat the critical behavior of a system undergoing second
order phase transition and has been shown to be a powerful method to obtain their critical indexes\cite{2}.

In an alternative way to the RG approach, C.J.Thompson\cite{3} used a heuristic method (of the dimensions) as a way
to obtain the correlation length critical index ($\nu$), which governs the critical behavior of a system in the
neighborhood of its critical point. Starting from Landau-Ginzburg-Wilson hamiltonian or free energy,he got a closed
form relation for $\nu(d)$ \cite{3}, where $d$ is the spatial dimension. It is argued that the critical behavior of
that $\Phi^4$-field theory is within the same class of universality as that of the Ising Model.

One of the present authors\cite{4} applied Thompson's method to study diffusion limited chemical reaction ${\bf A+A
\to 0}$ (inert product). The results obtained in that work\cite{4} agree with the exact results of Peliti\cite{5} who
renormalized term by term given by the interaction diagramms in the perturbation theory.

Nassif and Silva\cite{6} proposed an action to describe diffusion limited chemical reactions belonging to various classes of universality. This action was treated through Thompson's approach and can encompass the cases
of reactions like ${\bf A+B \to 0}$ and  ${\bf A+A \to 0}$ within the same formalism. Just at the upper critical
dimensions of ${\bf A+B \to 0}$ ($d_c=4$) and  ${\bf A+A \to 0}$ ($d_c=2$) reactions, the present authors found
universal logarithmic corrections for the mean field behavior.

Thompson's renormalization group method has been applied to obtain the correlation length critical exponent of the
 Random Field Ising Model by Aharony,Imry and Ma\cite{7} and by one of the present authors\cite{8}. His method was
also used to evaluate the correlation length critical exponent of the N-vector Model\cite{9}. Yang - Lee Edge
Singularity Critical Exponents\cite{10} has been also studied by this method. In short we have been exploring the
various possibilities of the Thompson's method of dimensions\cite{4}~\cite{6}~\cite{8}~\cite{9}~\cite{10}~\cite{11}~
\cite{12}~\cite{13}~\cite{14}~\cite{15}~\cite{16}~\cite{17}. As we can see, for instance, by considering these various
possibilities of the method, we were able to obtain the universal logarithmic behavior for the coupling parameters of
various models at their respective upper critical dimensions [4,6,8-16]. We also have shown how this method behaves
when applied to $QED_4$\cite{17} and we have obtained the logarithmic behavior on scale of energy for coupling
$\alpha$ (for $d=4$).

The aim of the present work is to describe firstly the QCD-vacuum behavior by considering a strong classical analogy
of such a vacuum with a paramagnetic material in the presence of an external magnetic field \cite{18}. To do that, in Section 3, we
will use a simple action in the form of that of Landau, i.e., without temperature and phase transition,
where the magnetization due to the presence of an external magnetic field $H$ is thought of as being a color scalar field
of virtual gluons. Such a cloud of virtual gluons are induced in vacuum because, in the investigation of the internal
structure of nucleons, high energy scales must be also considered. By applying Thompson's method (T.M) to such an
action, it will be possible to extract an ``effective electric  permissivity" $\epsilon<1$, an ``effective magnetic
permeability" $\mu>1$ and also an ``effective susceptibility" $\chi>0$, which depends on logarithmic of energy scale
$u$ used to investigate the hadronic matter. Just in order to obey Lorentz invariance, we take into account the simple
Lorentz condition for vacuum, i.e., $\mu\epsilon=1$ ($c^2=1$)\cite{18}. Our investigation provides an analogy
between the energy of the QCD-vacuum and the corresponding energy of magnetic dipoles of a paramagnetic material
being ligned up by the action of a magnetic field. Due to this fact, in Section 2, we will verify that QCD-vacuum at
high energies behaves as if it were predominantly a kind of ``color paramagnetism" for gluons with spin $1$, that is to say
the bosonic behavior of QCD-vacuum in such a ``paramagnetic regime ($\mu>1$)" for higher energies supplants completely
the fermionic contribution for vacuum due to ``diamagnetic regime ($\mu<1$)", and thus we will get the asymptotic
freedom in QCD as a consequence of this anti-screening effect (``vacuum paramagnetism")\cite{18}.

In Section 4, we will study the contribution of quantum fluctuations for the field energy density. This leads to an
interaction energy ($\Delta mc^2$) as an increment in the field energy and with logarithmic behavior on energy scale,
allowing us to obtain a $\beta-$ function to be compared with the well-known $\beta-$ function of QCD at one loop
level. Besides this, in contrast to the asymptotic freedom for high energies governed by those quantum
fluctuations, in Section 5, we will study quarks confinement regime for low energies (large distances) governed by
another new quantum contribution, which leads to a new increment in the field energy associated with the confinement
energy of quarks. The value of the well-known string constant is computed analytically and it is found to be in agreement with
the experiments.

\section{$QCD$-Lagrangian, color charges, gluons and the ``paramagnetism of color fields": Asymptotic freedom}

\subsection{$QCD$-Lagrangian}

Quantum Chromodynamics~(QCD)\cite{19}, the modern theory of the strong interactions\cite{20}~\cite{21} is a
non-abelian field theory. In 1973, Gross and Wilczek\cite{22} and independently Politzer\cite{23} have shown that
certain classes of non-abelian fields theories exhibit asymptotic freedom, which is a necessary condition for a theory that
describes the strong interactions. These seminal papers~\cite{19}~\cite{22}~\cite{23} opened the route to the birth
of QCD.

 In a not very accurate picture, QCD can be considered as an expanded version of QED. In QCD we have six fermionic
fields for representing the various quark flavors, in contraposition to a single fermionic field of QED. Besides the
asymptotic freedom exhibited at the ultraviolet limit, a theory of the strong interactions must also display quarks
confinement at the infrared limit.

 Whereas in QED there is just one kind of charge, QCD has three kinds of charge, labeled by ``color" (red, green and
blue)\cite{20}. The color charges are conserved in all physical processes. There are also photon-like massless
particles called color gluons that respond in appropriate ways to the presence of color charge. This mechanism has
some similarity with the ways that photons respond to electric charge in QED, except the non-abelian character of the
theory.

 Let us write the QCD-Lagrangian density, namely:

  \begin{equation}
L=\Sigma_j \overline\psi_j(i\gamma_{\mu}D^{\mu}- m_j)\psi_j - \frac{1}{4}G^a_{\mu\nu}
G_a^{\mu\nu},
  \end{equation}
where $D^{\mu}=\partial^{\mu}+\frac{1}{2}ig\lambda_a A^{\mu}_a$ , and $G^{\mu\nu}_a=\partial^{\mu}A^{\nu}_a-\partial^{\nu}A^{\mu}_a-gf_{abc}A^{\mu}_b
A^{\nu}_c$\cite{19} 

 In (1) above, $m_j$ and $\psi_j$ are the mass and quantum field of the quark of $j^{th}$ ``flavor", and $A$ is the
gluon field, $\mu$ and $\nu$ being the space-time indexes. $a$, $b$ and $c$ are color indexes. The coefficients $f$
(structure constants) and $\lambda_a$ guarantee $SU(3)$ color symmetry. $g$ is the coupling constant.

 The gluon part of (1) contains both a kinetic term 
 $L_{kin}=-\frac{1}{4}(\partial_{\mu}A_{a\nu}-\partial_{\nu}A_{a\mu})(\partial_{\mu}A_a^{\nu} -
 \partial_{\nu}A_a^{\mu})$ and an interaction term $L_{int.}= \frac{1}{2}g
 f_{abc}(\partial_{\mu}A_{a\nu}-\partial_{\nu}A_{a\mu})A_b^{\mu}A_c^{\nu}
-\frac{1}{4}g^2 f_{abc} f_{a^{\prime} b^{\prime} c^{\prime}} A_{b\mu}A_{c\nu} A_{b^{\prime}}^{\mu}
 A_{c^{\prime}}^{\nu}$. The form of the kinetic term is the same form as the photon term of the well-known
QED-Lagrangian. Thus exchange of gluons gives rise to forces similar to the Coulomb interaction but acting on
particles with color instead of charges. However, gluons carry color themselves (unlike photons that don't carry any
charge), leading to the interaction term ($L_{int.}$) between gluons themselves, and this is the situation that
makes QCD an asymptotically free theory. 

\subsection{Color charges and color fields (gluons)} 

 It is well-known the energy stored in an electric field according to the classical theory, mamely $U_{cl.}=\int_{V_3}
E_{cl.}^2 dV_3$, such integration performed in a $3D$ space-like volume. In a previous paper (see ref.[17]),
where we have considered $QED$ at high energies, quantum fluctuations due to vacuum polarization affect the energy
through a squared quantum contribution of the field ($\overline{\Delta E^2}$), since the linear quantum contribution
term $\Delta E$ averages out to zero for long time. So we have obtained $\overline{E^2}=\overline{E_{cl.}^2}+
\overline{\Delta E^2}$~\cite{17}, where the bars means averaging over a sufficiently long time at the scale of
fluctuations. Therefore, we were interested in the quantum process,namely the absorption and emission of virtual
photons,leading to the quantum correction in the field $E$, i.e., $\Delta E_{rms}=[(\overline\Delta
{E^2})]^{\frac{1}{2}}$~\cite{17}, where the index $rms$ means root mean square. We have thought that such a correction
is different of zero only in the presence of the fermionic field due to its purely quantum origin. This led us to
propose the relation $\Delta E_{rms}^2=\xi^2\psi_{rms}^2$~\cite{17},where we have considered
$\psi_{rms}^2=\left<[\overline\psi\psi]\right>_r=\frac{1}{2\pi^2 r^3}$ (see 11 in ref. [17]).  $\psi_{rms}^2$
corresponds to the mean squared fermionic field on the variable of $r$ scale, and $\xi$ is a proportionality
constant. Such relations allowed us to obtain $\Delta E_{rms}\propto\frac{1}{r^{\frac{3}{2}}}$ in QED for quantum
contribution of the field\cite{23}~\cite{24}~\cite{25} at high energy. It must be compared with the inverse square
Gauss law of the classical contribution, and so we realize that it leads to a logarithmic correction on $r$ scale for
energy of the field\cite{17}~\cite{25}.

 As QCD introduces color charges and color fields, since gluons carry color charges (unlike photons that don't carry
any charge)\cite{19},we could extend the reasoning above in order to treat QCD by considering a general ``color
electric field", namely:

 \begin{equation}
\overline{E_a^2}=\overline{E_{cl.a}^2} + \overline{\Delta E_{F.a}^2} + \overline{\Delta E_{B.a}^2}, 
 \end{equation} 
where $\overline{E_{cl.a}^2}$ is the classical contribution for the color field with a ``mode" $a$. $\overline
{\Delta E_{F.a}^2}$ is a fermionic contribution for the color field, which is similar to that of QED ($\Delta
E_{rms}=[(\overline{\Delta{E^2}})]^{\frac{1}{2}}$~\cite{17}), however QED has no color. $\overline{\Delta E_{B.a}^2}$
is a quantum contribution for the color field,which does not have any analogy with QED. Such a contribution is due to
quantum fluctuations of color fields in the presence of bosons (gluons) since they carry color themselves, i.e., 
it is a bosonic contribution for the color field. We will see that such a new quantum contribution has origin
exclusively from QCD-vacuum behavior, which leads to the anti-screening effect and thus makes QCD an asymptotically
free theory.

 Actually (2) supplies a total energy density $u = u_{cl.} + u_{F} + u_{B}$, $u_{cl}$ being the classical
contribution for energy density. $u_{F}$ and $u_{B}$ are the fermionic and bosonic contributions respectively. We will
see that $u_{B}$ has a changed signal with respect to $u_{F}$, which leads to the anti-screening effect and the
asymptotic freedom of QCD,in opposition to the screening effect of QED.

Now we assume that a heuristic approach used by Thompson\cite{3} to study critical phenomena can be applied to the
lagrangian (1). The first prescription of Thompson\cite{3}~\cite{17} is basically a scale argument with dimensional
analysis for average values on scales. It states that:

``When we consider the integral of the Lagrangian (1) in a coherence volume $L^d$ for $d$-dimensions, the modulus of
each integrated term of it is separately of the order of unity\cite{3}".

 This method by using its three prescriptions was firstly applied by Thompson\cite{3} to the Landau-Ginzburg-Wilson
free energy or Hamiltonian, obtaining critical exponents within the same universality class of the Ising model. As the
present model does not have any kind of phase transistion or spontaneous breakdown of symmetry, it is only necessary
to use the first prescription of Thompson. So by applying such prescription to the knetic fermionic term of
Lagrangian (1), we write

\begin{equation}
\left|\int_r [\overline\psi_j(i\gamma_{\mu}\partial^{\mu})\psi_j]_r dV_4\right|\sim 1,
\end{equation} 
where $dV_4\sim r^3 dr$ 

 We can observe that the dimension of \'~$\gamma_{\mu}\partial^{\mu}$\`~($[\gamma_{\mu}\partial^{\mu}]_r$) which
appears in the integral (3) is the same as $[\partial^{\mu}]_r = r^{-1}$. This is because we are thinking only about
a dimensional analysis in (3) for \'~$\gamma_{\mu}\partial^{\mu}$\`~. So, for this case, we can naturally neglect the
spinorial aspect of the field and just consider the \'~first derivative $\partial^{\mu}$\`~, which defines the
fermions (quarks) regarding to the scaling dimensional analysis, namely $[\partial^{\mu}]_r = r^{-1}$.

It is interesting to note that the integral above leads immediately to a kind of scaling dimensional analysis where
the dimensional value of certain quantity $[\overline{\psi_j}\psi_j]$ inside the integral is taken out of its
integrand as a mean value in a coherent hyper-volume of scale $L^4$, where $dV_4\sim r^3 dr$. Thus, from (3) we extract
the following scaling behavior, namely:

\begin{equation}
\left<[\overline{\psi_j}\psi_j]\right>_r \equiv [\overline{\psi_j}\psi_j]_r\sim\frac{1}{r^3}.
\end{equation}

In analogous way to that heuristic reasoning used for QED, by considering the fermionic contribution of condensate
$\left<\overline\psi\psi\right>$ over quantum fluctuations of field $E$, namely $\overline{\Delta E^2}\propto\left
<[\overline\Psi\Psi]\right>_r\propto\frac{1}{r^3}$~\cite{17}, then, for QCD, we have a fermionic contribution for quark
condensate, which also contributes for quantum fluctuations of the color field $E_a$ through $\Delta E_{F.a}$, namely:

\begin{equation} 
\overline{\Delta E_{F.a}^2}\propto\left<[\overline\Psi_j\Psi_j]\right>_r\propto\frac{1}{r^3}. 
\end{equation}

\subsection{``Paramagnetism of color fields" in QCD-vacuum: Asymptotic freedom}

Let us first recapitulate some properties of ordinary polarizable media for classical theory. In a polarizable
medium,the potential energy of two static test charges $q$ and $Q$ is $U_{el.}(r)=\frac{qQ}{4\pi\epsilon r}$, where
$r$ is the distance between the two charges, $\epsilon$ being the dieletric constant which takes the value
$\epsilon_0=1$ in vacuum. Ordinarily, the polarizability of the medium causes a screening of the interaction between the test
static charges, so that $\epsilon>1$. On the other hand, anti-screening corresponds to $\epsilon<1$.

A relativistic quantum field theory has a vacuum which presents a strong classical analogy with the ordinary
polarizable medium, however it just differs from an ordinary polarizable medium on a very important aspect: it is
relativistically invariant. This means that, if we set the speed of light $c=1$, the magnetic permeability $\mu$
is related to the dieletric constant (electric permissivity) $\epsilon$ by

 \begin{equation}
\mu\epsilon=1.
\end{equation}

The implication of Lorentz invariance in QCD is very important for theories about confinement of quarks and
gluons\cite{26}. Such a relationship (6) does not exist for an ordinary or classical polarizable medium.

In order to obey Lorentz invariance shown in (6), we can conclude that ordinary screening means $\mu<1$ (diamagnetism), and ordinary
anti-screening means $\mu>1$ (analogous to paramagnetism of the Landau's classical theory). The magnetic
permeability $\mu$ is written in the following way:
  \begin{equation}
 \mu=1+4\pi\chi,
\end{equation}
where $\chi$ is the magnetic susceptibility. QCD-vacuum has classical analogy to the paramagnetic medium\cite{18}.
We will see that the increase of energy scale $u$ for investigating the hadronic matter leads to the increase of ``effective
susceptibility" of QCD-vacuum $\chi_{eff.}=\chi(u)(>0)$ to be determined in the next section. This leads to an
increase of the ``effective magnetic permeability" of QCD-vacuum, namely $\mu_{eff}=1+4\pi\chi(u)$.

By considering a paramagnetic medium with a volume $V$ and an uniform magnetization $M$ in the presence of the field
$H$, thus we have the following energy:

\begin{equation}
E = E_{paramagnetic}= -\frac{1}{2} 4\pi MHV = -\frac{1}{2} 4\pi \chi H^2 V,
\end{equation}
where $M=\chi H$.

In spite of there is not Lorentz invariance in ordinary media, a paramagnetic medium still realizes a strong
classical analogy to QCD-vacuum in the sense that we could think that such vacuum is a medium with spin effect of
color charges\cite{18} related to virtual gluons with bosonic spin ($s=1$) like photons. So in this case, we have
a direct classical analogy to magnetization $M$ due to fermions ($s=\pm 1/2$), which leads us to think about a kind
of ``color magnetization $M_a$" for QCD-vacuum as being a ``color paramagnetic medium" in the presence of a ``color
magnetic field $H_a$". Following such an analogy to QCD-vacuum, we can write

\begin{equation}
M_a=\chi_{eff} H_a,
\end{equation}
where $a$ is just a color mode that we select for convenience, and $\chi_{eff}=\chi(u)$ is the ``effective
paramagnetic susceptibility" for QCD-vacuum, having dependence on the energy scale $u$.

 In (8), paramagnetism manifests itself through the minus sign in front of the right-hand side. This has an
analogy to QCD, where the vacuum energy is decreased in the presence of a color magnetic field\cite{18}. So (8)
can be written in the following way for representation of the ``color paramagnetic energy" of QCD-vacuum, namely:

\begin{equation}
E=E_{vac,QCD} = E_{color-paramagnetic}= -\frac{1}{2} 4\pi M_aH_aV = -\frac{1}{2} 4\pi \chi(u) H_a^2 V,
\end{equation}
where $V$ is a kind of coherence volume inside which the color fields are greatly correlated, having an analogy to the
correlated spin effect\cite{18}.

The behavior of the increasing function $\chi_{eff}=\chi(u)$ will be shown in the next section.

\section{An effective Landau's Hamiltonian as a model for mimicing the ``color paramagnetism''}

The strong classical analogy between QCD-vacuum at a certain energy scale $u$ of investigation and a
paramagnetic medium with magnetization $M$ in a magnetic field $H$ motivates us to introduce an effective Landau's
Hamiltonian for representing the vacuum inside the hadronic matter as a paramagnetic medium in the presence of a
color magnetic field $H_a$. This simple model will be presented in this section.

It is well-known that a cloud of virtual gluons emerges in QCD-vacuum at high energies $u$,leading to the ``color
paramagnetism" (anti-screening), whereas, on the other hand, a cloud of virtual electron-positron pairs appears in
QED-vacuum at high energies, leading to the vacuum polarization. We have a ``dielectric vacuum" (screening) for QED.

The cloud of virtual gluons in QCD-vacuum are quanta of the color field induced by the probe used to investigate the
structure of the hadronic matter,which depends on its energy scale $u$. It depends also on the proper color magnetic
field $H_a$ that already exists inside the hadronic matter under investigation. Therefore,such a color field could be
thought of as being directly related to the color magnetization $M_a$ and also to the color magnetic field $H_a$,
since we have the relation $M_a=\chi(u)H_a$. So now let us think about such a color field as being a general scalar
field $\Phi_a$,namely:

\begin{equation}
\Phi_a=\Phi_a(r)=\Phi_a[M_a(r),H_a(r)]=\Phi_a[\chi(u),H_a(r)],
\end{equation}
where $M_a(r)=\chi(u)H_a(r)$. As the effective susceptibility $\chi_{eff.}=\chi(u)$ and the color magnetic field
$H_a$ are independent parameters, let us use for convenience the scalar color field $\Phi_a(r)$ in the form
$\Phi_a(r)=\Phi_a[\chi(u),H_a(r)]$. Here we think that the color magnetic field $H_a$ and color magnetization $M_a$
have dependence on $r$-coordinate inside the ``color paramagnetic medium" represented by the hadronic matter.

Due to the classical analogy to paramagnetism, let us introduce now the following ``effective Landau Hamiltonian"
for mimicing the ``color paramagnetism", namely:

\begin{equation}
F=\int_{L^d}[(\nabla\Phi_a)^2 + R(L)\Phi_a^2 + K(L)\Phi_a^4]d^dr,
\end{equation}
where, in this case, the coefficients $R(L)$ and $K(L)$ do not depend on temperature. 

The integration (12) extends over $d$-dimensional volume. Thompson's approach has three assumptions (ref.[3]). 
As we are not interested in phase transition\cite{3} in this model, we must use only the two first ones, namely:

(A) When the integral in (12) is taken over the coherence volume $L^d$ in $d$-dimensions, the three terms separately
in (12) are all of the order of unity.\cite{3}~\cite{17}

(B) In the specific case of (12), we just have to consider the parameter $K(L)$ to be finite in the limit
$L\rightarrow\infty$. This leads us to consider a mean field regime above a certain critical dimension $d_c$ where
the coefficient $K$ remains constant. In Landau's theory, we have $d_c=4$.\cite{3} 

By applying the assumption (A) in the first term of (12), we write

\begin{equation}
\int_{L^d} (\nabla\Phi_a)^2 d^dr\sim 1,
\end{equation}
where the parameter $L$ forms the basis of our dimensional argument and may be thought of as being a wavelength cut-off,
so that the mean value $\overline\Phi_a^2$ behaves as

\begin{equation}
\overline{\Phi_a^2}\sim L^{2-d}.
\end{equation}

For the second term in (12), we have

\begin{equation}
\int_{L^d} R(L)\Phi_a^2 d^dr\sim R(L)\overline{\Phi_a^2} L^d\sim 1.
\end{equation}

By inserting (14) into (15), we obtain

\begin{equation}
R(L)L^2\sim 1.
\end{equation}

For the third term in (12), we have

\begin{equation}
\int_{L^d} K(L)\Phi_a^4 d^dr\sim K(L)\overline{\Phi_a^4} L^d\sim 1,
\end{equation}
such that, from (14) and the assumption (B), so from (17), we obtain

\begin{equation}
 K(L)\sim\left\{
\begin{array}{ll}
 L^{d-4}:&\mbox{$d\leq 4$},\\\\
 1:&\mbox{$d\geq 4$.}
 \end{array}
\right.
\end{equation}

From (18), we observe that $d=d_c=4$ is a special dimension (an upper critical dimension) above which we have a mean field
behavior\cite{6}, that is to say the coupling parameter $K$ does not depend on scale $L$, so that $K$ is a constant parameter.
In other words, below $d=4$, fluctuations are very important for the problem, deviating from the mean field behavior, and
above $d=4$, the ``mean field" description\cite{6} is a good description for the problem. So $d=4$, which coincides with
the space-time dimension, corresponds exactly to a kind of border-line dimension for representing the QCD-vacuum as a Lorentz
invariant theory and also its classical analogy with paramagnetic media. Therefore, we must improve our approximations
in order to ``see" the logarithmic dependence on scale $L$ (ref.[17]) of the coupling $K(L)$ just at $d=d_c=4$, or
equivalently at the energy scale $u=L^{-1}$. Similar situation has also occurred when we treated diffusion limited
chemical reactions through Thompson's approach\cite{4}~\cite{6}~\cite{11}~\cite{12}~\cite{13}, displaying universal
logarithmic behavior at ``upper critical dimensions" for ``coupling constants" of those different models. Following
that improvement technique to ``see" such a logarithmic behavior, let us improve the calculation of (17) by taking the
quantity $\Phi_a^4$ inside the integral (17) and starting from the same scale form as that evaluated in (14), but now
displaying a dependence on $r$-variable of scale. So by taking inside the integral (17) the quantity
$[\Phi_a]_r^4=([\Phi_a]_r^2)^2= r^{4-2d}$ and also the $d$-dimensional volume of integration in the form
$d^dr=r^{d-1}dr$, we have

\begin{equation}
\int K(r) r^{4-2d}r^{d-1}dr=\int K(r) r^{3-d}dr\sim 1,
\end{equation}
where, just for $d=4$, exactly on the boder-line of mean field regime where $K$ is practically constant, we are
able to see now the refinement of the logarithmic dependence at length scale for $K$, namely $\int K r^{-1}dr\sim 1$,
which implies $K\sim [ln(r)]^{-1}$. So now, if we perform such a integration between the limits of scales $L$ and
$L_0$, by considering $L_0$ an upper cut-off of length, we write

\begin{equation}
\int_{L}^{L_0} K r^{-1} dr\sim 1,
\end{equation}
from where we obtain

\begin{equation}
K=K(L)\sim\frac{1}{ln(\frac{L_0}{L})}\sim K(u)\sim\frac{1}{ln(\frac{u}{u_0})},
\end{equation}
where the energy scales are $u=1/L$ and $u_0=1/L_0$, with $u>u_0$, since $u_0$ is a lower cut-off at the scale of
energy, i.e., it is an infrared limit.

 As we have obtained the logarithmic behavior of the coupling parameter $K$ just at $d=d_c=4$ for such a paramagnetic
medium, we can make an analogy with the $QCD_4$-vacuum by obtaining now the ``color scalar field amplitude" $\Phi_a^2$,
having a direct analogy with the equilibrium magnetization in the Landau picture, namely $M^2=-\tau r(L)/u(L)$\cite{3}. However, since
in the present model we do not have any spontaneous breakdown of symmetry like in the Landau picture\cite{3}, here we just consider the
coefficient $R(L)$ instead of ``$\tau r(L)$''\cite{3} which depends on temperature. Thus we obtain

\begin{equation}
\Phi_a^2 =-\frac{R}{K}
\end{equation}

As we are interested only in the behavior of $\Phi_a^2$ on the border-line at $d_c=4$ associated with the
space-time dimension, we insert (21) into (22) and so we find

\begin{equation}
\Phi_a^2(r)=-c_1R(r)~ln\left(\frac{u}{u_0}\right),
\end{equation}
where $c_1>0$ is a positive proportionality constant.

We can associate the amplitude of scalar field $\Phi_a^2(r)$ with a negative energy density $\rho(r)$ of a ``color
paramagnetic medium" ($QCD_4$-vacuum),having analogy to that negative energy density of a paramagnetic medium,namely
$-\frac{1}{2}4\pi\chi H^2$ (ref.[18]). However we must consider an ``effective susceptibility" $\chi_{eff.}=\chi(u)$
to represent QCD-vacuum,and also consider a ``color magnetic field" $H_a(r)$. Thus such an analogy leads us to write

\begin{equation}
\rho_{vac,QCD}(r)=\Phi_a^2(r)=-c_1R(r)~ln\left(\frac{u}{u_0}\right)\equiv-\frac{1}{2}4\pi\chi(u)H_a(r)^2,
\end{equation}
from where we can firstly extract $c_1\equiv 2\pi$ and $R(r)\equiv H_a^2(r)$, and thus we can rewrite (24) as follows:

\begin{equation}
\rho_{vac,QCD}(r)=\Phi_a^2(r)=-c_1R(r)~ln\left(\frac{u}{u_0}\right)\equiv -\frac{1}{2}4\pi H_a(r)^2~ln\left(\frac{u}{u_0}\right).
\end{equation}

By comparing the right side of (25) with the right side of (24), we can also extract the ``effective susceptibility",
as follows:

\begin{equation}
\chi_{eff}=\chi(u)=ln\left(\frac{u}{u_0}\right).
\end{equation}

From (26), it is interesting to observe that the effective susceptibility of $QCD_4$-vacuum increases logarithmicaly
with energy scale $u$. From (21) and (26), we can also notice that the parameter $K$ is $K(u)\sim\chi(u)^{-1}$, which
allows us to interpret such parameter as being related to a ``strength" of coupling $\alpha_S$ between quarks. So
we have $\alpha_S\sim K$. That is because, when $u\rightarrow u_0$ for the infrared limit, this implies
$\chi(u_0)\rightarrow 0$ (very weak ``paramagnetism"), which leads to $\alpha_S(u_0)\sim K(u_0)\rightarrow\infty$
(a much stronger coupling), namely we have a highly confined regime of quarks for low energies (infrared regime). On the other
hand, when $u\rightarrow\infty$ for the ultraviolet regime, this implies $\chi(u)\rightarrow\infty$ (``color paramagnetism"
becomes much more evident), which leads to $\alpha_S(u)\sim K(u)\rightarrow 0$ (a very weak coupling between quarks),
that is to say we have the well-known asymptotic freedom (high energies).

For sake of simplicity, if we take the color magnetic field $H_a$ practically uniform in (25), namely a
uniform energy density $\rho_{vac,QCD}$ (or $\Phi_a^2$), and by considering a coherence volume $V$, we simply obtain
the``color paramagnetic energy" $E$ as that given in (10), where $\chi_{eff}$ is now given in (26). So we finally write

\begin{equation}
 E_{vac,QCD}=\rho_{vac,QCD}V= \Phi_a^2 V= -\frac{1}{2}4\pi H_a^2~ln\left(\frac{u}{u_0}\right)V.
\end{equation}

The ``effective magnetic permeability" $\mu(u)=1+4\pi\chi(u)$ can be obtained by considering (26), namely:

\begin{equation}
\mu(u)=1+4\pi~ln\left(\frac{u}{u_0}\right). 
\end{equation}

In order to obtain the ``effective electric permissivity" or the dieletric constant $\epsilon(u)$ of QCD-vacuum, now
we must guarantee the Lorentz invariance by considering the relation (6) ($\mu\epsilon=1$). So in doing that and
considering (28), we find

\begin{equation}
\epsilon(u)=\frac{1}{1+4\pi~ln(\frac{u}{u_0})},
\end{equation}
where $u\geq u_0$. We have $\mu=\mu_0=1$ for $u=u_0$.

In QCD, we have an anti-screening such that the effective interaction between strong charges for higher energies is
$Q_{eff.S}^2=\epsilon q_S^2$, with $\epsilon<1$,that is $Q_{eff.S}<q_S$. As the strong interaction is directly related
to the strong coupling $\alpha_S$, we can also write it in the form $\alpha_{S}=\epsilon~\alpha_{0S}$. So by
considering (29), finally we can also write it as follows:

\begin{equation}
\frac{\alpha_S}{\alpha_{0S}}=\frac{1}{1+4\pi~ln(\frac{u}{u_0})},
\end{equation}
where we fix $\alpha_{0S}$ to be a large value, but finite for low energies. So (30) reveals to us the asymptotic
freedom behavior for $QCD_4$ at high energies because, if we fix $u_0$ and consider $u\rightarrow\infty$, the ratio
$\frac{\alpha_S}{\alpha_{0S}}\rightarrow 0$. This means that the strong coupling decreases when the energy scale
increases. However, only the bosonic contribution of gluons for QCD-vacuum was evaluated for dieletric
constant. In reality, there is a competition between the effects of bosonic (anti-screening) and fermionic (screening)
contributions, where the first one prevails. This subject will be treated in the next section.

\section{Contribution of quantum fluctuations for energy}

 Quantum fluctuations lead to an interaction energy ($\Delta mc^2$) as an increment in the field energy and with
logarithmic behavior at length or energy scale. We can represent both of the fermionic and bosonic contributions of
energy density $u_{F}$ and $u_{B}$ (see (2)) in the following conpact form:

\begin{equation}
u_T=u_{qF} + u_{qB}\equiv u_{cl}\left(\frac{r}{\overline\lambda_c}\right)=
\frac{1}{2}\frac{1}{4\pi}\alpha_s\hbar c \frac{1}{r^4}\left(\frac{r}{\overline\lambda_c}\right),
\end{equation}
where $u_{cl}$ represents the classical energy density. $u_T$ is the total contribution of quantum fluctuations for
energy density, such that there is certain superior cut-off wavelength $\overline\lambda_c$ below which
($r<\overline\lambda_c$) we have quantum behavior of energy density $u_T$ on $r$ scale, i.e., $u_T=u_q\propto 1/r^3$,
and equal or above which ($r\geq\overline\lambda_c$) we recover the well-known classical behavior of
$u_T$, namely $u_T=u_{cl.}\propto 1/r^4$. To be more accurate, we rewrite a general form of $u_T$ for two regimes, namely:

\begin{equation}
 u_T=\left\{
\begin{array}{ll}
  u_{q}=u_{qF}+u_{qB}=\frac{1}{2}\frac{1}{4\pi}\frac{\alpha_s\hbar c}{\overline\lambda_c r^3}:
 &\mbox{$r\leq\overline\lambda_{c0}$}\\\\
  u_{cl}=\frac{1}{2}\frac{1}{4\pi}\frac{\alpha_s\hbar c}{r^4}: &\mbox{$r\geq\overline\lambda_{c0}$},
 \end{array}
\right.
\end{equation}
where $\overline\lambda_{c0}=\hbar/m_0c$ is a sharp cut-off wavelength. As the quantum regime also presents the
bosonic contributions $u_B$, which leads to anti-screening in QCD, the mass $m_0$ must be considered as a dynamical
variable which exhibits fluctuations depending on the energy scale.

We are interested only in quantum regime for energy density $u_q$ of the field ($r<\overline\lambda_{c0}$). So we
want to obtain the interaction energy $\Delta E=\Delta m c^2$ in a certain coherence volume $V$. Then let us think
about a spherical volume $V$ and therefore we have the interaction energy in the differential form,namely:

\begin{equation}
dE=dmc^2=u_q 4\pi r^2dr=\left(\frac{1}{2}\frac{1}{4\pi}\frac{\alpha_s\hbar c}{\overline\lambda_c r^3}\right)4\pi r^2dr.
\end{equation}

We can also write (33) in the following way:

\begin{equation}
dE=dmc^2=\frac{1}{2}\frac{\alpha_s\hbar c}{\overline\lambda_c}\left(\frac{dr}{r}\right)=
-\frac{1}{2}\frac{\alpha_s\hbar c}{\overline\lambda_c}\left(\frac{du}{u}\right),
\end{equation}
where we have considered the energy scale $u$ such that $r=u^{-1}$ and $dr/r=udr=-du/u$. So by performing the
integration of (34), we write

\begin{equation}
\Delta mc^2=-\frac{1}{2}\int_{u_0}^{u}\frac{\alpha_s\hbar c}{\overline\lambda_c}\left(\frac{du}{u}\right)=
-\frac{1}{2}\int_{u_0}^{u} \alpha_s mc^2\left(\frac{du}{u}\right),
\end{equation}
where $\overline\lambda_c=\hbar/mc$ and $u>u_0$. Due to fluctuations, it is natural to think that the coupling
$\alpha_s$ and the mass $m$ vary rapidly with the energy scale $u$, so that we must take them off the integral (35) as
avarage values on scale $u$, and thus we find 

\begin{equation}
\Delta m = -\frac{1}{2}{\overline\alpha_s}~{\overline m}\int_{u_0}^{u}\frac{du}{u},
\end{equation}
where we define $(\alpha_s m)_{average}= {\overline\alpha_s}~{\overline m}$.

On the other hand, it is important to perceive that the increment on mass $\Delta m=(\int_Vu_qdV)/c^2$ due to
interactions in such a quantum regime is directly proportional to the increment on the coupling ($\Delta\alpha_s$),
since both the increments present logarithmic behavior on scale. Such reasoning was used before for QED\cite{17} and so now
by extending it to our problem, let us write

\begin{equation}
\frac{\Delta\alpha_s}{\overline\alpha_s}=\frac{\Delta m}{\overline m}
\end{equation}

By inserting (36) (for $\Delta m$) into (37) and performing the calculations, we obtain

\begin{equation}
\alpha=\alpha_{0s}-\frac{1}{2} (\overline\alpha_s)^2~ln(\frac{u}{u_0}),
\end{equation}
where $\Delta\alpha_s=\alpha_s-\alpha_{0s}$.

Now let us write $(\overline\alpha_s)^2$ in the following way:

\begin{equation}
(\overline\alpha_s)^2=\alpha_s\alpha_{ref},
\end{equation}
where $\alpha_{ref}$ is a certain reference coupling to be duly interpreted.

The equations (36),(37) and (39) define the variable parameters $\overline\alpha_s$,$\overline m$ and $\alpha_{ref}$.
 Only the parameters $\alpha_s$, $m$, $\Delta\alpha_s$ and $\Delta m$ are always real quantities, since they are physical
parameters.

By substituting (39) in (38), we obtain
\begin{equation}
\alpha_s=\alpha_s(u)=\frac{\alpha_{0s}}{1+\frac{\alpha_{ref}}{2}~ln(\frac{u}{u_0})}.
\end{equation}

From the general result (40), we can observe that the sign of $\alpha_{ref}$ can change by controlling the
predominance of anti-screening or screening. In the case of $\alpha_{ref}>0$, then we have $\alpha_s\rightarrow 0$ for
$u\rightarrow\infty$, which leads to the asymptotic freedom behavior of QCD connected to anti-screening. On the other
hand, if $\alpha_{ref}<0$, we have the well-known Landau singularity\cite{17}, namely a finite value of the
energy scale $u_L$ such that $\alpha(u_L)\rightarrow\infty$\cite{17}. For illuminating ideas about the Landau pole,
see refs.\cite{27}~\cite{28}. This is the case of the increasing of the coupling with the
increasing of the energy scale associated with screening for QED. This opposite case ($\alpha_{ref}<0$) leads to an
imaginary value for $\overline\alpha_s$ (see (39)), which means a non-asymptotically free theory like QED.

The result (40) also implies the following differential equation:

\begin{equation}
u\frac{d\alpha_s}{du}= -\left(\frac{\alpha_{ref}}{\alpha_{0s}}\right)\alpha_s^2. 
\end{equation}

 By performing the integration of (41) above in the limits $u_0$ and $u$ and their respective couplings
$\alpha_s(u_0)$ and $\alpha_s(u)$, we obtain (40).

This differential equation (41) must be compared with the well-known $\beta$-function for QCD\cite{18}~\cite{29}~\cite{30}
~\cite{31} when evaluated at one loop level, namely:

\begin{equation}
 \left[u\frac{d\alpha_s}{du}\right]_{QCD}=-\left(\frac{33 - 2n_F}{3\pi}\right)\alpha_s^2.
\end{equation}

It would be worth to make some comparison of the results of the equation (41) or (40) with some experimental
evaluation of the strong coupling as a function of momentum (energy) of the probe. In a plot of reference [20],
it is possible to get an estimate for $\alpha_{0s}$, namely:

\begin{equation}
\alpha_{0s}=\alpha_s(u_0=1Gev)\cong 0.43.
\end{equation}

Taking in consideration (43) and comparing (41) with (42), we have

\begin{equation}
\alpha_{ref}=\alpha_{0s}\frac{21}{3\pi}\cong 0.96.
\end{equation}

  For obtaining $\alpha_{ref}$ in (44), we also have considered $n_F=6$ as the number of quark flavors. By using
(40) and the fact that $\alpha_{ref}\equiv\alpha_s(u=u_{ref})\cong 0.96$, thus $u_{ref}$ can be determined.
 So taking into account the previous results, we get

\begin{equation}
u_{ref}\cong 316 Mev.
\end{equation}

This reference value is practically the quark constituent mass of the nucleon since we consider that each valence
quark carries out one third of the nucleon mass.

\section{Quarks Confinement}

 In contrast to the asymptotic freedom, which governs the ultraviolet (short distances) behavior of the theory, we
also intend to look for its infrared behavior. We propose that such achievement can be done by taking into
consideration the role of a scalar field in addition to the other quantum contributions for the energy density.
 Whereas $u_{qF}$ and $u_{qB}$ behaves as $1/r^3$,we have $u_{confine}\equiv u_{qc}\propto 1/r^2$. So we write the
total field energy density as follows:

\begin{equation}
u_T=u_{cl}+u_{qF}+u_{qB}+u_{qc},
\end{equation}
where $u_{qc}$ is supplied by the scale relation (14) for $d=4$, namely
$u_{qc}\propto\left<{\Phi^2_a}\right>\propto 1/r^2$. For very large distances (much lower energies),
$u_{qc}$ prevails whereas the terms $u_{cl}\propto 1/r^4$, $u_{qF}\propto 1/r^3$ and $u_{qB}\propto 1/r^3$ go rapidly
to zero. That is the reason why that last term in (46) governs quarks confinement.

 As we can relate $\Phi_{ia}$ to a gluon field with a dressed gluon mode $i$ and a single color indexed by $a$,
let us write

\begin{equation}
u_{qc}\propto\left<\Phi_{ia}\Phi_{ia}\right>\propto\frac{1}{r^2},
\end{equation}
where $i=1,2...8$, with the $3$ colors indexed by ``$a$".

 Let us compare $u_{qc}$ with $u_{cl}$ in a similar way made for $u_{qF}$ and $u_{qB}$ in (32), and so we write

\begin{equation}
u_{qc}=u_{cl}\frac{r^2}{\overline\lambda_c^2}=\frac{1}{2}\frac{1}{4\pi}\alpha_s\hbar c
\frac{1}{\overline\lambda_c^2 r^2},
\end{equation}
where $u_{cl}=\alpha_s\hbar c/8\pi r^4$. $\overline\lambda_c=\hbar/m_qc$, $m_q$ being the quark constituent mass.
 Therefore, by inserting $\overline\lambda_c$ into (48), we obtain

\begin{equation}
u_{qc}=\frac{1}{2}\frac{1}{4\pi}\frac{\alpha_s m_q^2c^3}{\hbar}\frac{1}{r^2}. 
\end{equation}

 Now we can obtain the confinement total energy $\Delta E_{confine}=\Delta m_{confine}c^2$ by performing the
integration of (49) over a spherical volume and also by considering the total of $8$ gluons with $3$ colors. So we write

\begin{equation}
\Delta m_{confine}c^2=24\int u_{qc} dV_3=
24\int_0^r\left(\frac{1}{2}\frac{1}{4\pi}\frac{\alpha_s m_q^2c^3}{\hbar}\frac{1}{r^2}\right)4\pi r^2 dr. 
\end{equation}

  From (50), we finally obtain

\begin{equation}
\Delta m_{confine}c^2=12\overline\alpha_s\frac{m_q^2c^3}{\hbar}\int_0^r dr,
\end{equation}
where $\overline\alpha_s=\alpha_s(m_{nucleon}c^2)=\alpha_s(940Mev)$.

 Since there are $3$ quarks inside the nucleon, having each one the constituent mass $m_q=(1/3)m_{nucleon}$, and also $3$ 
pairs of linked quarks\cite{32}, thus the confinement mean energy per quark pair $\Delta\epsilon$ leads us to consider
from (51) that

\begin{equation}
\Delta\epsilon=\frac{1}{3}(\Delta m_{confine}c^2)=4\alpha_s(940Mev)\frac{m_q^2c^3}{\hbar}\int_0^r dr=k_sr,
\end{equation}
where we think there are $3$ pairs of strongly coupled quarks, so that each pair is internally linked by a string constant
$k_s=4\alpha_s(940Mev)m_q^2c^3/\hbar$.

 We can estimate the value of $k_s$. Taking into account (43) and (44), we can calculate $\alpha_s(940Mev)$ from
(40), and so we obtain $\alpha_s(u=940Mev)\cong 0.443$. We also obtain $m_q^2c^3/\hbar=m_q^2c^4/\hbar c\cong 0.498GeV/fm$, where we 
have considered that each valence quark with constituent mass $m_q$ carries out one third of the nucleon mass. Finally we 
estimate $k_s\cong 0.882$ Gev/fm. This result is in agreement with experiments.\cite{33}

 We must emphasize that the non-perturbative treatment for infrared regime in QCD is generally based on
lattice gauge theory\cite{34}, leading to numerical approaches. Therefore, quarks confinement cannot be treated
analytically by perturbative methods. In view of this fact, we perceive a great advantage of the present approach 
that was able to obtain analytically by a simple way the value of the string constant.\\

{\noindent\bf  Acknowledgedments}

We are grateful to Prof. Holger Gies from Heidelberg University for clarifying the problem of Landau pole in QED and
also for interesting suggestions in applying the method to non-perturbative problems like the quarks confinement in QCD.
The first author thanks Prof. J. A. Helayel-Neto for interesting discussions.

\end{document}